\documentclass{iau} 
\usepackage{graphicx}
\usepackage{caption}
\usepackage{subcaption}

\title[The Hubble Catalog of Variables] 
{The Hubble Catalog of Variables}

\author[P. Gavras et al.]   
{P.~Gavras$^1$, A.~Z.~Bonanos$^1$, I.~Bellas-Velidis$^1$, V.~Charmandaris$^{1,2}$, I.~Georgantopoulos$^1$, D.~Hatzidimitriou$^{3,1}$, G.~Kakaletris$^4$, A.~Karampelas$^1$, N.~Laskaris$^4$, D.~J.~Lennon$^5$, M.~I.~Moretti$^{1,6}$, E.~Pouliasis$^{1,3}$, K.~Sokolovsky$^1$, Z.~T.~Spetsieri$^{1,3}$, K.~Tsinganos$^{1,3}$, B.~C.~Whitmore$^7$, M.~Yang$^1$ }

\affiliation{$^1$IAASARS, National Observatory of Athens, Penteli, Greece \\
$^2$Department of Physics, University of Crete, Heraklion, Greece\\
$^3$Department of Astrophysics, Astronomy \& Mechanics, Faculty of Physics, National \& Kapodistrian  University of Athens, Athens, Greece \\
$^4$Athena Research \& Innovation Center, Athens, Greece\\
$^5$European Space Astronomy Centre, ESA, Villanueva de la Canada, Madrid, Spain\\
$^6$INAF Osservatorio Astronomico di Capodimonte, Naples, Italy\\
$^7$ Space Telescope Science Institute, Baltimore, USA\\}

\pubyear{2016}
\volume{325}  
\jname{Astroinformatics 2016}
\editors{M. Brescia, S.G. Djorgovski, E. Feigelson, \\G. Longo \& S. Cavuoti, eds.}

\begin{document}

\maketitle

\begin{abstract}
The Hubble Catalog of Variables (HCV) is a 3 year ESA funded project that aims to develop a set of algorithms to identify variables among the sources included in the Hubble Source Catalog (HSC) and produce the HCV. We will process all HSC sources with more than a predefined number of measurements in a single filter/instrument combination and compute a range of lightcurve features to determine the variability status of each source. At the end of the project, the first release of the Hubble Catalog of Variables will be made available at the Mikulski Archive for Space Telescopes (MAST) and the ESA Science Archives. The variability detection pipeline will be implemented at the Space Telescope Science Institute (STScI) so that updated versions of the HCV may be created following the future releases of the HSC.
\end{abstract}

\firstsection 
\section{Introduction}
The Hubble Catalog of Variables is a project aiming to automatically detect photometric variability, among the sources contained in the Hubble Source Catalog (\cite[Whitmore et al. 2016]{Whitmore}). The current version of the HSC contains more than 90 million sources observed over the 26 years of  \textit{Hubble Space Telescope} (HST). The catalog is very diverse, as it includes observations obtained with more than 100 filter/instrument combinations (see contribution by Steve Lubow, this volume). Our plan is to process all HSC sources that have more than a predefined number of observations in one or more filter/instrument combinations. The process involves the identification and rejection of bad measurements, improving the photometry by determining and applying local zero-point corrections, calculating lightcurve features and creating a list of candidate variables. After processing the HSC, the list of candidate variables will automatically be validated to filter out artifacts and generate a list of true variables. According to the requirements of the project, the pipeline should not take longer than a couple of months to run. At the end of our project, the pipeline and the catalog (HCV) will be incorporated in the Mikulski Archive for Space Telescopes (MAST) maintained by STScI. ESA Science Archives will host the catalog as well.

\section{Methodology}
Our goal is to exploit the data contained in the HSC, therefore direct image comparison techniques are not applicable. Moreover, since we are not aiming to detect a specific type of variable, a generic period search will not be useful.
In order to identify variables within the HSC, the HCV pipeline uses lightcurve features which quantify the variability of a lightcurve. We call these features Variability Indexes (VIs). Our pipeline has a library of 20 VIs, which we classify as scatter-based and correlation-based indexes. The first category takes into account the scatter of the points of the lightcurve, while the second uses the correlation between the points. Table 1 shows the list of variability indexes we have implemented in our pipeline. The first column shows the name of the VI, while the next three indicate the use of the magnitude errors, the order  and the time of observations, respectively.
 A detailed description of the variability indexes and their effectiveness can be found in \cite[Sokolovsky et al. (2017)]{Sokolovsky}. We have selected a large number of different VIs as they may highlight different variability patterns, thus by using all these indexes, our pipeline is able to detect many different types of variables.
\begin{table}
  \begin{center}
  \caption{The list of variability indexes calculated by the HCV pipeline.}
  \label{tab1}
 {\scriptsize
  \begin{tabular}{lcccc}\hline 
{\bf Index} & {\bf Errors} & {\bf Order} & {\bf Time} 
  \\ \hline
\multicolumn{4}{c}{Scatter-based indexes}  \\
reduced $\chi^2$ statistic - $\chi^2_{red}$  & $\surd$ \\
weighted standard deviation - $\sigma_w$ &$\surd$  \\
median absolute deviation - MAD  \\
interquartile range - IQR  \\
robust median statistic - RoMS& $\surd$ \\ 
normalized excess variance - $\sigma^2_{NXS}$ & $\surd$ \\ 
normalized peak-to-peak amplitude - u& $\surd$ \\
\multicolumn{4}{c}{Correlation-based indexes}\\
Stetson's I index& $\surd$ &$\surd$ & $\surd$ \\
Stetson's J index&  $\surd$ &$\surd$ & $\surd$ \\
time-weighted Stetson's J(time) & $\surd$ &$\surd$ & $\surd$ \\
clipped Stetson's J(clipped) & $\surd$ &$\surd$ & $\surd$ \\
Stetson's L index & $\surd$ &$\surd$ & $\surd$ \\
time-weighted Stetson's L(time) & $\surd$ &$\surd$ & $\surd$ \\
clipped Stetson's L(clipped)& $\surd$ &$\surd$ & $\surd$ \\
consecutive same sign deviations - CSSD & &$\surd$ \\
excursions - $E_x$ & $\surd$ &$\surd$ & $\surd$ \\
autocorrelation - $l_1$&&$\surd$\\
inverse von Neumann ratio - 1/$\eta$&&$\surd$\\
excess Abbe value $\mathcal{E}_A$&  &$\surd$ & $\surd$ \\
$S_B$ statistic & $\surd$ &$\surd$  \\ \hline
  \end{tabular}
  }
 \end{center}
\vspace{1mm}
\end{table}

\begin{figure}[h]
\begin{subfigure}{0.48\textwidth}
\begin{center}
 \includegraphics[clip,width=\textwidth]{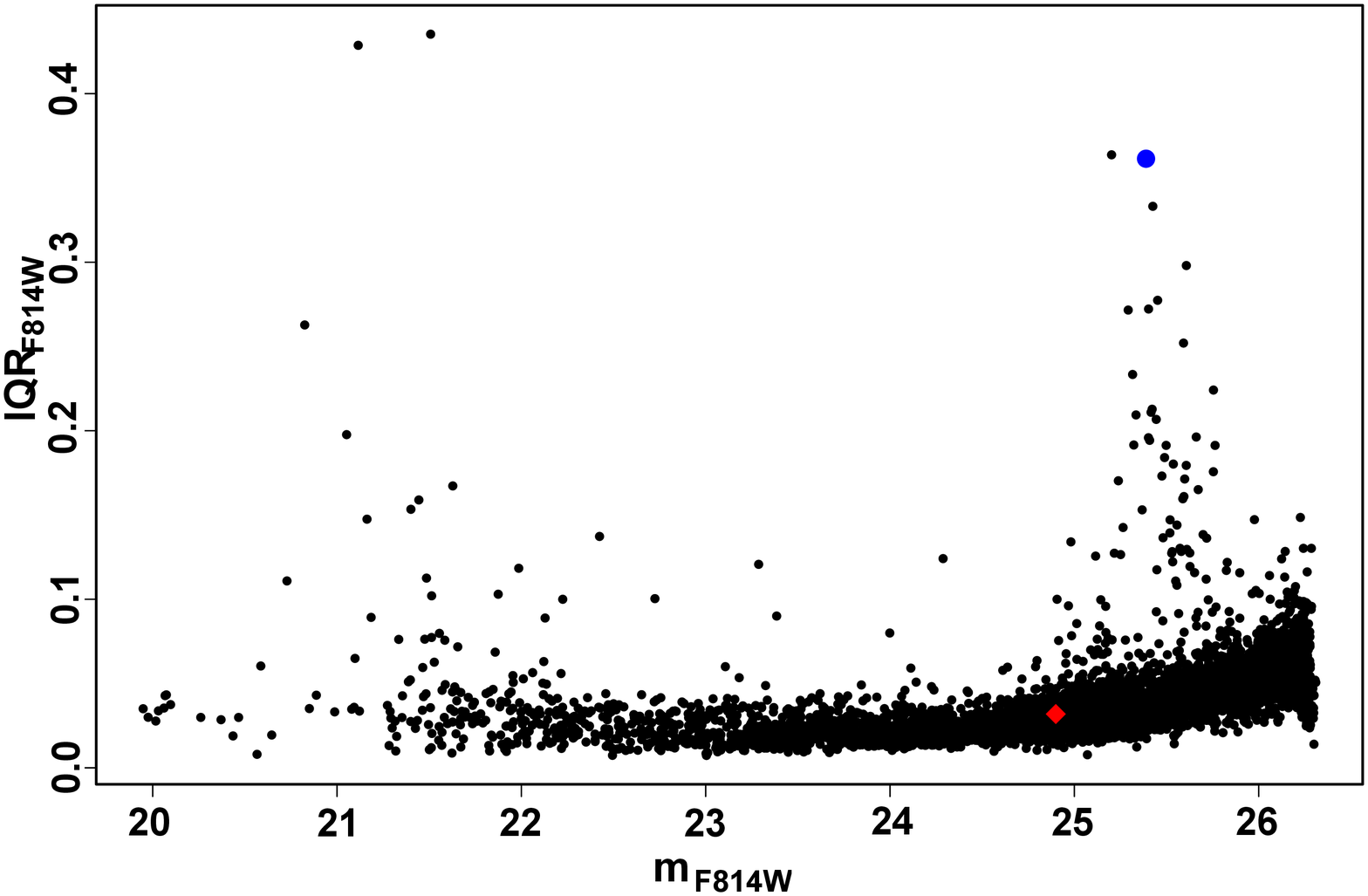} 
   \end{center}
  \end{subfigure}%
  \hfill
\begin{subfigure}{.48\textwidth}
\begin{center}
 \includegraphics[width=\textwidth]{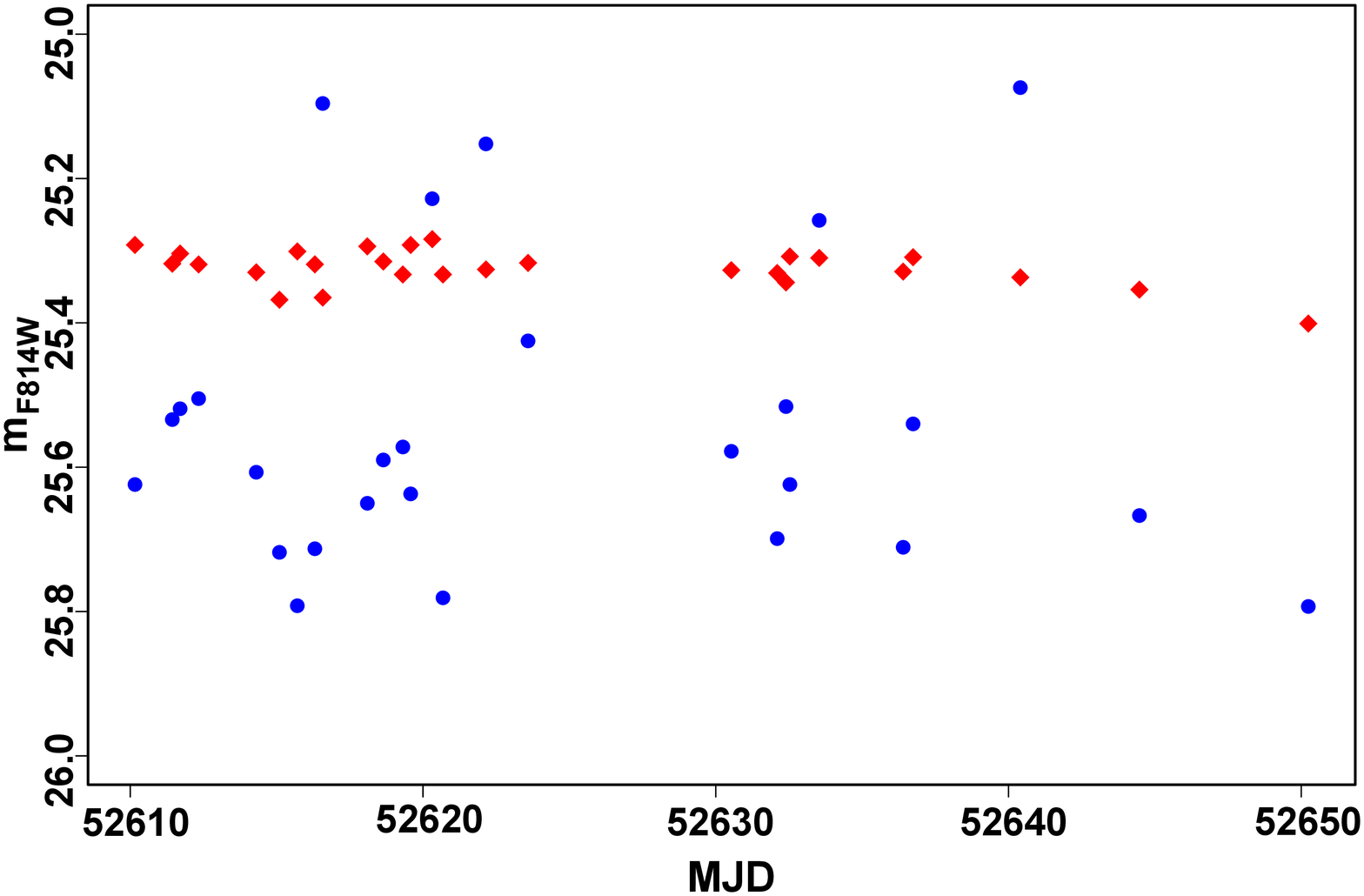} 
\end{center}
\end{subfigure}%
\caption{ \textit{Left}: Magnitude vs. IQR diagram for a field in M31. We mark 2 sources, one with a high value (blue) and one with a low value (red) of IQR. \textit{Right}: Lightcurves of the selected sources.}
\label{fig:fig1}
\end{figure}

\begin{figure}[h]
\begin{subfigure}{0.48\textwidth}
\begin{center}
  \includegraphics[clip,width=0.8\textwidth]{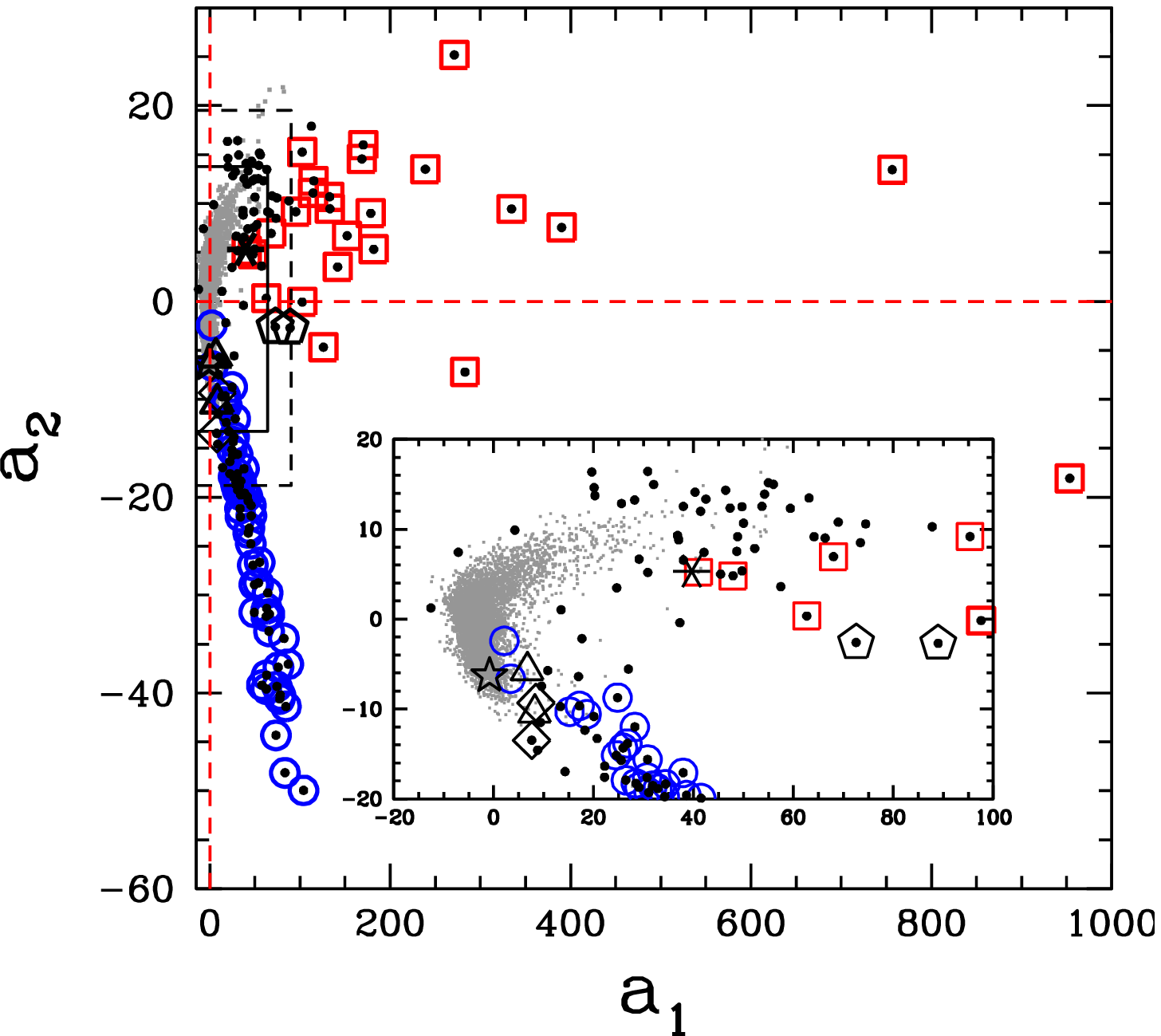} 
\end{center}
\end{subfigure}
\hfill
\begin{subfigure}{.48\textwidth}
\begin{center}
 \includegraphics[clip,width=0.8\textwidth]{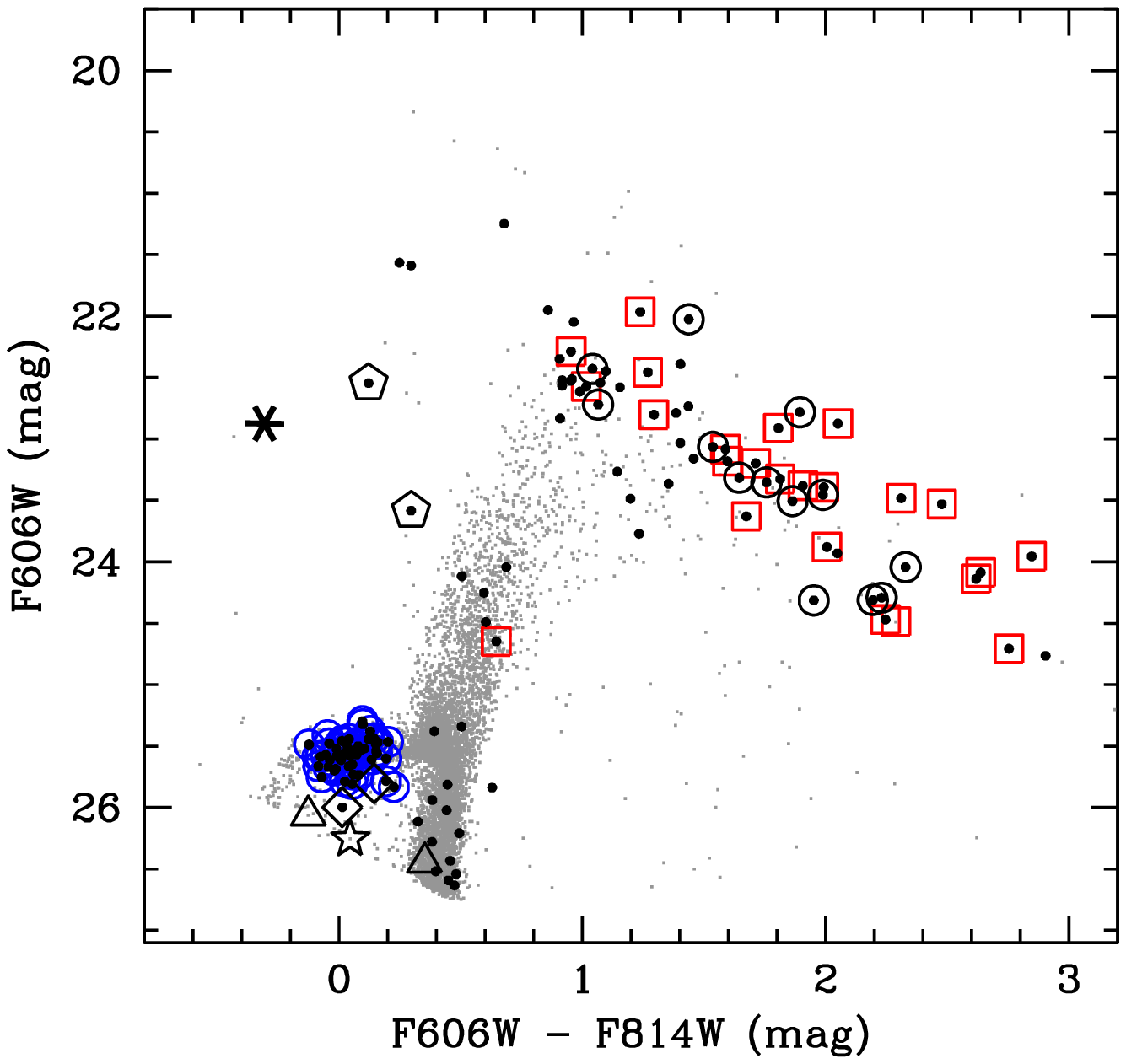} 
\end{center}
\end{subfigure}%
\caption{ \textit{Left}: a1, a2 plot for the M31 Halo11 sources. 
\textit{Right}: CMD diagram for the Halo 11 sources. The PCA candidates are marked with black points. Known variables are also highlighted.}
\label{fig:fig2}
\end{figure}

We assume that a variable source will have a value in at least one VI  that is significantly different from the value expected for a constant source of similar brightness observed in the same field.
 As the HSC contains millions of sources and a small percentage are variable, we expect to have a "sea of constant sources" and few variables that stand out in a magnitude vs. variability index diagram. In Figure \ref{fig:fig1} we present such a diagram for the interquartile range (IQR) variability index and 2 sources have been selected as an example.  The blue source clearly stands out from the locus of constant stars on the plot. This source is variable, as illustrated by its lightcurve in Figure \ref{fig:fig1}. On the other hand, the red source which lies in the region of the diagram with low IQR values, is constant.\\
The next step is to reduce dimensionality by applying principal components analysis (PCA) on the normalized variability indexes.
This analysis showed that PC1 contains about 60\% of the variance in the data.\\
Finally, we consider as candidate variables those sources that do not cluster with the main population in an admixture coefficients diagram like the one presented in Figure \ref{fig:fig2}, where the admixture coefficients a1 and a2 are plotted.
\begin{figure}[h]
\begin{center}
 \includegraphics[clip,width=0.48\textwidth]{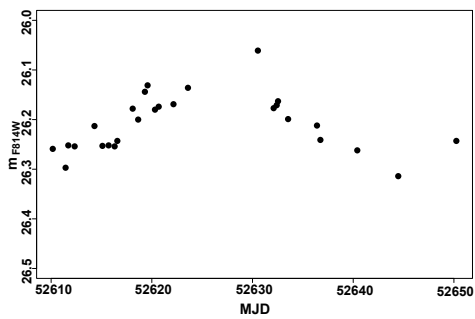} 
 \caption{ HSC lightcurve of a newly identified variable candidate.}
   \label{fig:fig3}
\end{center}%
\end{figure}
 
\section{Implementation}

The HCV is implemented as a distributed pipeline system that employs state of the art technology to assure its extensibility and scalability objectives. It utilises, among other technologies, a distributed computing engine (Apache Spark) to manage the distribution of work over the computational and memory resources and a high throughput distributed filesystem (HDFS) to host and serve data, compute intermediate results and outputs. The entire system resides on a multi-host virtualised infrastructure.

\section{Application to M31}

We applied our algorithm to a field located along the M31 minor axis, at 11 kpc from the center (M31 Halo11 hereafter), trying to recover all known variables and detect new ones. M31 Halo11 was observed by HST with the Advanced Camera for Surveys (ACS; \cite[Avila et al., 2016]{Avila}) in two filters, F606W \& F814W, with mean number of observations 27 and 32, respectively. Using our method we were able to recover $\sim$90\% of the known variables existing in the HSC and identify about 50 new candidates (Moretti et al., in prep). In Figure \ref{fig:fig2}, our PCA candidates are marked with black points, while the various types of known variables from \cite[Brown et al., (2004)]{Brown} are presented with different shapes and colors. 
In particular, LPVs/semiregulars are shown with red empty squares, RR Lyrae with blue empty circles, RR Lyrae candidates are black empty diamonds, eclipsing binaries are presented as black empty triangles, $\delta$ Scuti stars are black empty stars, anomalous Cepheids are shown with black empty pentagons and a post-AGB star with a black asterisk.
Furthermore, the color-magnitude diagram is presented using the same color coding. Figure \ref{fig:fig3} shows an example of a lightcurve of one of our newly identified variable candidates.

\section{Summary}
We highlighted the basic aspects of the Hubble Catalog of Variables project. In HCV we will process all the astronomical objects in the Hubble Source Catalog by calculating a wide variety of lightcurve features (i.e. variability indexes). We will also apply a PCA analysis to identify the variable sources using a multi-dimensional optimization approach. We demonstrated an application of the method to the M31 Halo11 field where we recovered $\sim$90\% of the known variables and identified about 50 new candidates, which are currently under investigation.
\section*{ACKNOWLEDGEMENTS}
We acknowledge financial support by the European Space Agency (ESA) under the ‘Hubble Catalog of Variables’ programme, contract no. 4000112940.

\end{document}